\documentclass{article}

\usepackage{arxiv}

\usepackage{graphicx}
\usepackage{multirow}
\usepackage{amsmath,amssymb,amsfonts}
\usepackage{amsthm}
\usepackage{mathrsfs}
\usepackage[title]{appendix}
\usepackage{xcolor}
\usepackage{textcomp}
\usepackage{manyfoot}
\usepackage{booktabs}
\usepackage{algorithm}
\usepackage{algorithmicx}
\usepackage{algpseudocode}
\usepackage{listings}

\usepackage[english]{babel}
\usepackage[utf8]{inputenc}

\usepackage{hyperref}
\hypersetup{ colorlinks,
linkcolor=blue,
filecolor=blue,
urlcolor=blue,
citecolor=blue }

\usepackage{amsmath}
\usepackage{xcolor}

\title{New Coevolution Dynamic as an Optimization Strategy in Group Problem Solving}

\author{
  Francis F.~Franco \\
   Department of Physics\\
   Federal University of Jata\'i (UFJ)\\
   Jata\'i, GO, Brazil, 75801-615\\
   \texttt{francis.franco@educa.go.gov.br} \\
   \And
 Paulo F.~Gomes \\
   Department of Physics\\
   Federal University of Jata\'i (UFJ)\\
   Jata\'i, GO, Brazil, 75801-615\\
  \texttt{paulofreitasgomes@ufj.edu.br} \\
}

\begin{document}
\maketitle


\begin{abstract}
Coevolution on social models couples the time evolution of the network with the time evolution of the states of the agents. This paper presents a new coevolution dynamic allowing more than one rewiring on the network. We explore how this coevolution can be employed as an optimization strategy for problem-solving capability of task-forces. We use an agent-based model to study how this new coevolution dynamic can help a group of agents whose task is to find the global maxima of NK fitness landscapes. Each agent can replace more than one neighbor, and this quantity is a tunable parameter in the model. These rewirings is a way for the agent to obtain information from individuals that were not previously part of its neighborhood. Our results showed that this tunable coevolution can indeed produce gain on the computational cost under certain circunstances. At high average degree network and difficult landscape, the effect is complex. If the agent has a low fitness, 3 or 4 rewirings can bring some improvement.
\end{abstract}



\maketitle

\section{Introduction}

In society, it is common to face problems that require collaboration with other people, from everyday challenges to complex tasks, such as group projects at work. In this context, the search for more effective problem-solving strategies becomes a topic of great interest. Social interactions are inherently complex such that, in recent years, numerous models of social behavior have emerged to explore this intricacy \cite{Castellano2009,Perc2010,Jusup2022} using the framework of Network Science \cite{Barabasi2016}. The nodes of the network represent individuals (or agents) with states defined by the social model while the connections represent the interactions between different individuals. The states of the individuals evolve following the set of rules from the dynamic social model in study. The network itself can also have a time evolution: links can be destroyed or created. 

Presently, a prominent aspect under exploration within dynamic social processes on complex networks is coevolution: the evolution of the agent states is coupled to the evolution of the network \cite{Gross2008,Marceau2010,Choi2023}. This coevolution has been studied in different social models \cite{Zimmermann2005,Qin2015,Raducha2018,Reia2019a,Saeedian2020,Reia2020JSM,Gomes2022} and Evolutionary Games Theory \cite{Szolnoki2008,Szolnoki2009}. The rationale behind this dynamics is that co-evolutionary, or adaptive, network models offer a more accurate portrayal of real world systems compared to static or evolving networks. Empirical networks display both network dynamics (the evolution of network topology) and node state dynamics. Therefore, coevolution can be conceptualized as two simultaneous processes – link rewiring and state dynamics – mutually influencing each other

The central idea of coevolution in social dynamics is the agent uses the social information (states of the agents) of his influence neighborhood (set of neighbors) to remove one neighbor and add a new one. And then, considering this new influence neighborhood, his social state will evolve in time accordingly. This mechanism can make a difference in cooperative processes, where one studies the performance of a set of individuals solving a given task \cite{Maier1960,Lazer2007,Reia2019x}. Each individual not only use his influence network, but also changes it if he thinks one neighbor may not give a proper contribution to solve the problem. This is specially important in imitative learning, where one tries to learn from an expert's experience and behavior \cite{Rendell2010,Zare2024}. 

How to modify the influence network is an important point to achieve success. Analyzing a group performance in problem-solving can be intricate. Initially, one might assume that a larger number of individuals working towards the same objective would expedite reaching a solution. However, interpersonal interactions (such as everyone engaging in discussion) can lead to suboptimal outcomes, deviating from the desired solution \cite{Fontanari2014Plos,Capraro2024}.

In this work we study the effect of a coevolution mechanism on a cooperative process problem. To address this question, we use an agent-based model where the agents can perform individual trial-and-test searches to probe a fitness landscape (exploration) or
imitate a model agent – the best performing agent in their influence neighborhood at the trial (exploitation). The coevolution mechanism gives a chance for the agent to change his influence neighborhood on the exploitation case. We consider a scenario where the agents are fixed at the nodes of a random network and can interact with each other if they are connected. In addition, before each interaction the agent has a chance to change $L$ agents within his influence neighborhood.
This set of rules creates a time-dependent and adaptive network. 

This work is organized as follows: In Section \ref{nklkjoiuwer} we present a brief discussion about the NK model, in section \ref{computacoskjer} we define the computational cost, in section \ref{slkjeiuiuiuiui} we define how the agents are updated, in sections \ref{lkjmnmnmnmnmn} and \ref{eiuerncbcbdhd} we define the proposed coevolution mechanism and in section \ref{gfgfgfgfgfgfgfgf} we described the random network used to connect the agents.Section \ref{sec3} details some required methods to analyze the results. Section \ref{resultskj} is dedicated to present and discuss the results and section \ref{slkjeroiuejww} is dedicated to our final considerations.

\section{Model}

\subsection{NK Fitness Landscapes} \label{nklkjoiuwer}

The NK model was introduced by Kauffman and Levin  as a fitness model for adaptive evolution processes such as walking in rugged landscapes \cite{Kauffman1987,Kauffman1989}. This model has been extensively used in different studies of cooperative process and imitative learning \cite{Kauffman1989b,Kauffman1993,Fontanari2016a,Barkoczi2016,Ganco2017}. It is well suited for example for the imperfect version of the imititive learning process \cite{Yao2022}, when one agent copies only part of the solution from an expert. 

The system is composed of $M$ agents described by a binary string (or vector) of $N$ bits $\mathcal{X} = ( x_1,x_2,..., x_N)$ with $x_i \in \{0,1\}$. In total there are $2^N$ different vectors $\mathcal{X}$. Each string has a fitness $f(\mathcal{X})$, limited in the interval $0.0 < f(\mathcal{X}) < 1.0$, calculated as \cite{Fontanari2016EPL}:
\begin{equation}
f(\mathcal{X}) = \frac{1}{N} \sum_{i=1}^N \Phi_i (K,\mathcal{X}), \label{wkjeriuwlkj}
\end{equation}
where $\Phi_i (\mathcal{X})$ is the contribution of component $i$ to the fitness of string $\mathcal{X}$. It depends on the state $x_i$ as well as on the states of the $K$ right neighbors of $i$, i.e.,
\begin{equation}
\Phi_i (K,\mathcal{X}) = \Phi_i (x_i,x_{i+1},x_{i+2},...,x_{i+K}), \label{lkjweoeiruhkjh}
\end{equation}
with the arithmetic in the subscripts done modulo $N$ \cite{Baumann2024}. The functions $\Phi_i$ are $N$ distinct real-valued functions on $\left\lbrace 0,1 \right\rbrace^{K+1}$. As usually done, we assign to each $\Phi_i$ a uniformly distributed random number in the unit interval, so that $f(\mathcal{X}) \in (0,1)$ \cite{Reia2017EcoCom}.

The function $f(\mathcal{X})$ has a global maximum $f(\mathcal{X}_G) = F$ at the string $\mathcal{X}_G$. The number of local maxima of the function $f(\mathcal{X})$ is zero at $K = 0$ landscape and increases at $K> 0$. Therefore, at $K=0$, the fitness difference $f(\mathcal{X}_G)-f(\mathcal{X})$ is proportional to the Hamming distance $d_H (\mathcal{X}_G , \mathcal{X})$ between the vectors \cite{Gomes2019}. The Hamming distance gives the number of different bits between the two vectors. For example, consider $\mathcal{X}_1 = (010101)$ and $\mathcal{X}_2 = (000101)$. The hamming distance is $d_H ( \mathcal{X}_1 , \mathcal{X}_2 ) = 1$. At $K > 0$, the landscape becomes more complicated because the function $f$ has additional local maxima beside the global one. Thus the difference $f(\mathcal{X}_G) - f(\mathcal{X})$ is no longer proportional to the hamming distance $d_H$. The average number of local maxima increases with $K$ until the limiting case of $K = N-1$, in which the landscape becomes completely random, known as the Random Energy model \cite{Derrida1981}.

\subsection{Computational cost} \label{computacoskjer}

The time evolution of the system is measured in Monte Carlo steps (MCS). The evolution gets to a halt when one agent reaches the maximum fitness $F$ of the landscape. Each agent changes its fitness $f(\mathcal{X})$ when its string $\mathcal{X}$ is updated (described in next section). So, all agents try to increase their fitness during the time evolution. The dynamics begins at $t=0$ with the selection of one agent for a string update. One Monte Carlo step (MCS) is the update of $M$ random agents, in sequence.

The efficiency of the search is measured by the time $t_g$ in MCS required by the first agent who finds the global maximum $F$, in other words, when its string becomes $\mathcal{X}_G$ after a number of updates. A small $t_g$ means more efficient search: the goal was achieved in less time. So, the lower the better. The normalized computational cost is then defined as \cite{Reia2019a}:
\begin{equation}
C \equiv M t_g / 2^N. \label{compcostlk}
\end{equation} 
This normalization ensures $C$ can be compared between systems of different sizes $M$ and landscapes $N$.

It is important to compare this computational cost with the one from the system considering $M$ independet agents. This means that each agent does not interact with any other. In this case, the agents are placed on the nodes of a completely disconnected network, containing $M$ components. This case is used as an reference and has an analytic result \cite{Fontanari2015EPJB}:
\begin{equation} 
    \left\langle C_I \right\rangle = \frac{M}{2^N \left[ 1- \left( \lambda_N \right)^M   \right] }, \label{indsearch}
\end{equation}
where $\lambda_N$ depends on the value of $N$. $\left\langle \cdot \cdot \cdot \right\rangle$ means an average over independend random samples of the time evolution. $C>C_I$ means the interaction between the agents creates more problems than solution, slowing down the quest of the agents for the global maximum. Otherwise, if $C<C_I$, the interaction helped the agents making one of them find the global maximum in less time.

\subsection{Strings update} \label{slkjeiuiuiuiui}

Each agent has its string $\mathcal{X}$ updated when it is selected for analysis. This is how each agent tries to improve its fitness $f(\mathcal{X})$, to eventually reaches the global maximum $F$, which happens when $\mathcal{X} = \mathcal{X}_G$. Some basic definitions we will use throughout the text are:
\begin{itemize}
\item Target: selected agent for the analysis, with string $\mathcal{X}_t$ and fitness $\phi_t = f(\mathcal{X}_t)$.
\item Influence neighborhood: set of agents who are neighbors of the target $\mathcal{X}_t$.
\item Model $\mathcal{X}_m$: agent from this neighborhood with the highest fitness, with fitness $\phi_m = f(\mathcal{X}_m)$.
\item Worst neighbor $\mathcal{X}_s$: agent on the neighborhood with the smallest fitness. We write its fitness as $\phi_s = f(\mathcal{X}_s)$.
\end{itemize}
The target $\mathcal{X}_t$ may be the model $\mathcal{X}_m$ or the one with the smallest fitness, $\mathcal{X}_s$, so that we have $\phi_s \leq \phi_t \leq \phi_m$.

 The target can implement one of the two strategies to perform the update of its string $\mathcal{X}_t$: exploitation or exploration \cite{Baumann2024}. Exploitation implies the target tries to learn the solution from a neighbour with a higher fitness. This is implemented as a copy, with probability $p$: the target copies one bit of the model string $\mathcal{X}_m$ \cite{Gomes2019}. Otherwise, exploration means the target aims to innovate a new solution with a better fitness by itself. It is a mutation on the string of the target, and it should happen with probability $1-p$. However if the target has no neighbors or if it is the model of its influence neighborhood, $\mathcal{X}_t = \mathcal{X}_m$, the mutation will happen for sure. If it has no neighbors, it does not have anyone to copy. And if it is already the model, there is no point in copy another agent with a smaller (worst) fitness.

Both processes change the string of the target agent. The mutation is achieved by a flip of a random bit of the string. The copy process involves more steps: the different bits bewteen the target and the model strings, $\mathcal{X}_t$ and $\mathcal{X}_m$, are identified. Then, one of them is randomly selected and flipped on the target string $\mathcal{X}_t$. This is the imperfect copy on the imitation learning context \cite{Yao2022}. As a result, the Hamming distance $d_H (\mathcal{X}_t,\mathcal{X}_m)$ between the two strings decreases. This means the target agent increases his fitness, as $f (\mathcal{X}_m) > f(\mathcal{X}_t)$. If $K=0$, it means the target gets closer to the global maximum. On the other hand, if $K>0$, the model may be a local maxima and the target $\mathcal{X}_t$ can get further away from the global maximum $\mathcal{X}_G$: $d_H(\mathcal{X}_t,\mathcal{X}_G)$ increases while $d_H(\mathcal{X}_t,\mathcal{X}_m)$ decreases. The target can get trapped on the suboptimal solution (local maximum) \cite{Lazer2007}. $K=0$ is the easy landscape and increasing $K$ increases the difficulty of the landscape. The most difficult landscape is $K=N-1$, where there is no point on copy because the landscape is completely random.

\begin{figure*} [t]
\centering
\includegraphics[width=6.1 in]{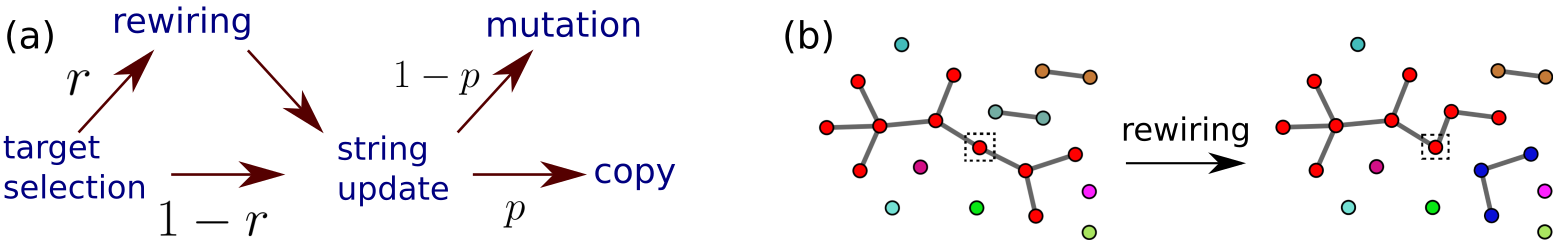}
\caption{Schematic illustration of the implemented model. (a) Illustration of one iteration. It starts with the selection of the target agent. $r$ is calculated through Eq. \ref{coevo}. With probability $r$ the network can suffer a rewiring. The next step is the string update, where a copy happens with probability $p$ and a mutation with probability $1-p$. (b) Illustration of the rewiring process. The target agent (indicated by a dashed black square) loses $L$ neighbors and gets new $L$ ones. The case $L=1$ is illustrated in this picture: the target loses the neighbor on the right and gets acquainted with the agent right above him. The colors of the nodes indicate the components.}
\label{fig0}
\end{figure*}

\begin{figure} 
\centering
\includegraphics[width=2.2 in]{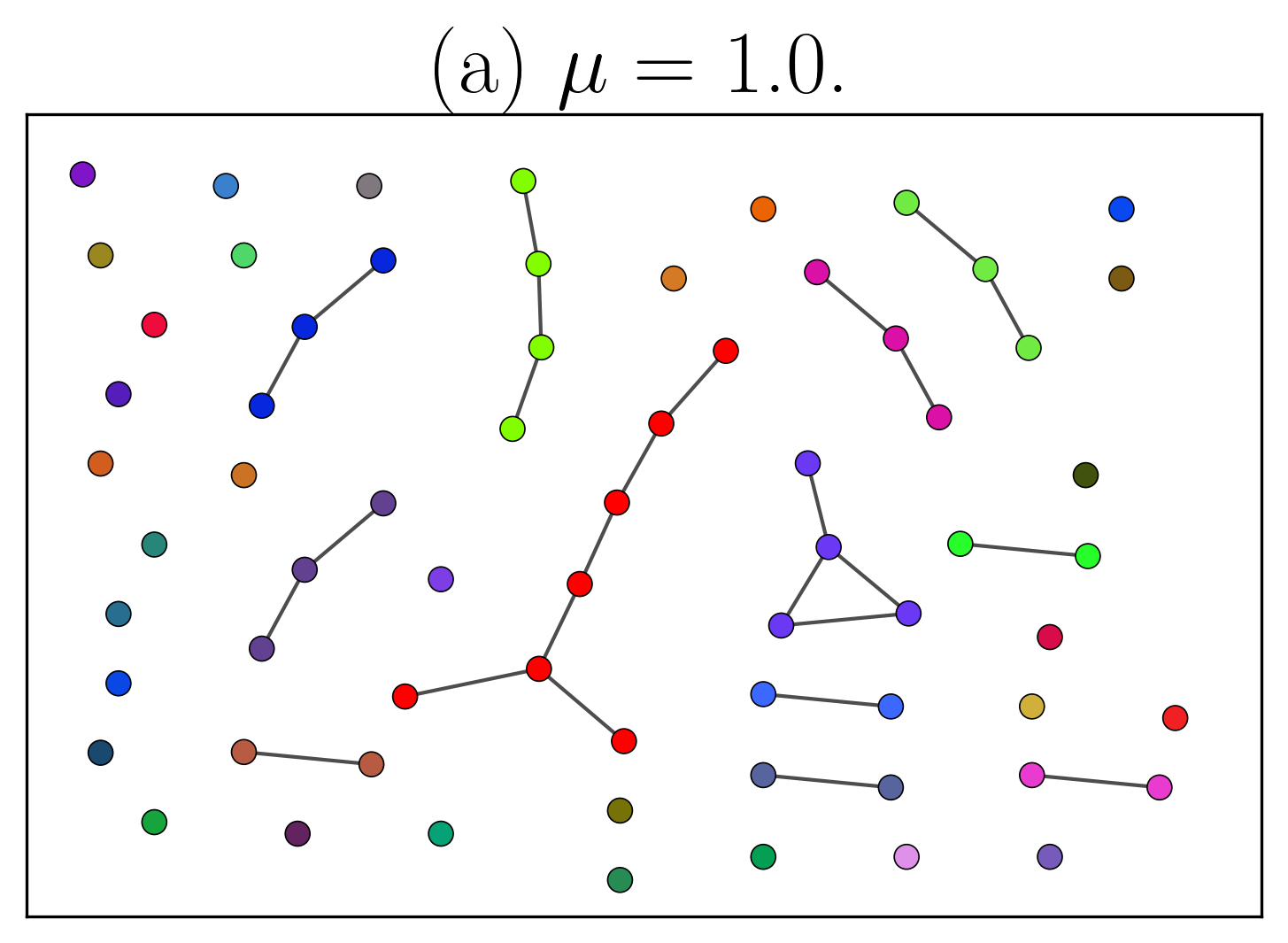}
\includegraphics[width=2.2 in]{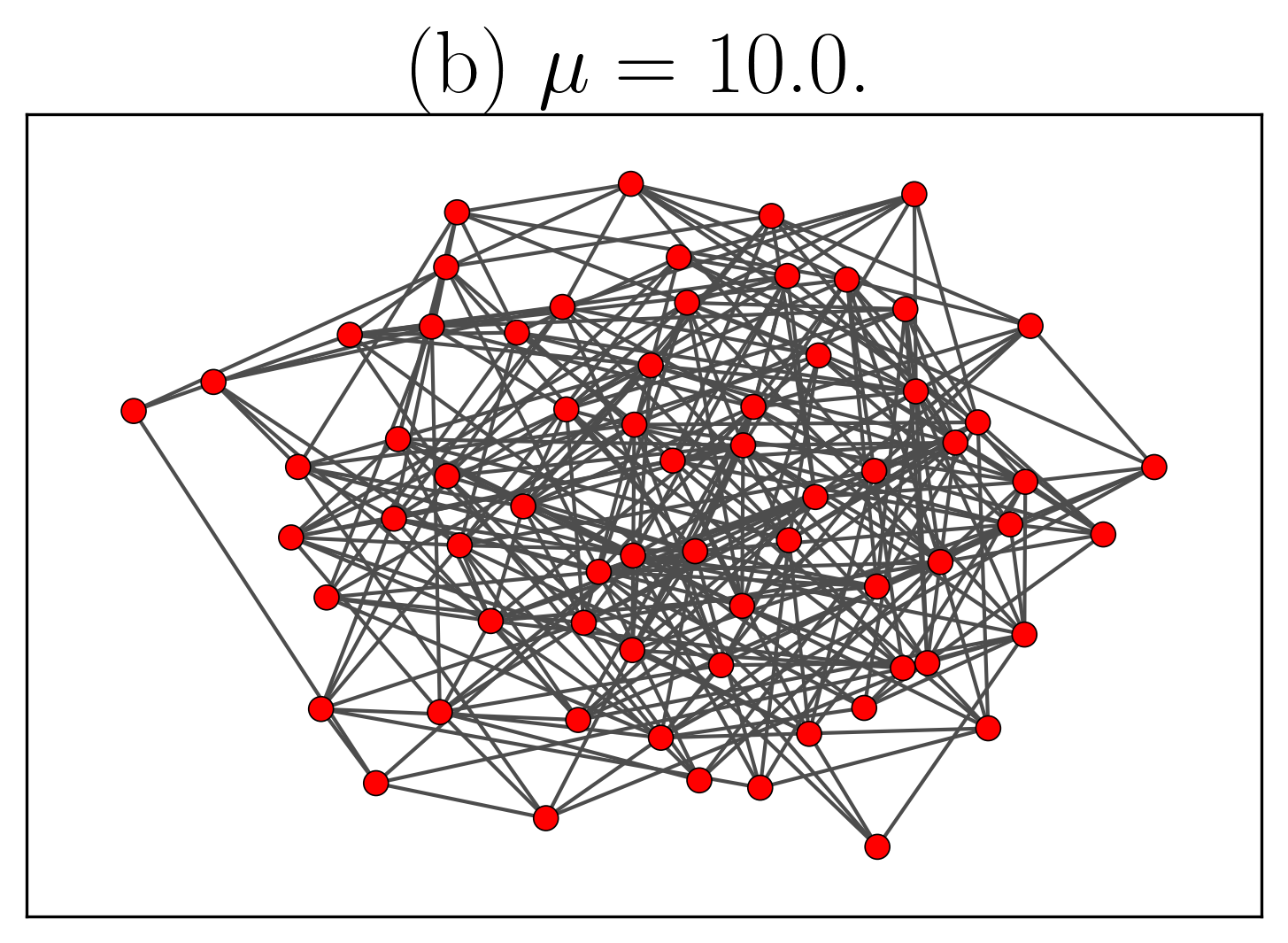}
\caption{Snaphots of instances of the random network used in this work. System size: $M=67$ agents. Each component is represented in a different color and the largest one is in painted in red. (a) Low average degree scenario: $\mu = 1.0$. The network is not percolated. (b) High average degree scenario: $\mu = 10.0$. In this scenario the network is percolated.}
\label{figXsdff}
\end{figure}

\subsection{Coevolution mechanism} \label{lkjmnmnmnmnmn}

The general idea of coevolution is to apply a rewiring before each string update: removing a random neighbor and adding a new one to the target agent. However, we can let the number of destroyed and created connections be greater than one. Therefore, before each string update we remove $L<k_i$ random neighbors and add $L$ new random neighbors to the target agent. The parameter $L$ is an input to our model, it is the control parameter of the coevolution process. The static network scenario is recovered with $L=0$.

The rewiring itself is the temporal evolution of the network. To be considered as coevolution, this network evolution must be coupled to the evolution of the agent string. The constant rewiring (always happening) is not coevolution, although it is a network evolution \cite{Gomes2019}. As the rewiring itself is a random process, the coupling of the network evolution with the agent states, in our model, will be achieved by the probability of rewiring $r$: it must be calculated considering the fitness $\phi$ of the agents. The rewiring process happens before the string update, as illustrated on Fig. \ref{fig0}(a). Another illustration of one example of the rewiring itself (with $L=1$) is shown Fig. \ref{fig0}(b).

\subsection{Rewiring probability} \label{eiuerncbcbdhd}

The goal of each agent is to increase its fitness $f(\mathcal{X})$ (given by Eq. \ref{wkjeriuwlkj}) until one of them, eventually, reaches the global maximum fitness $F$ of the landscape. So, having neighbors with better fitness if beneficial. One can propose that the probability $r$ should be larger when most of the neighbors have smaller fitness, so one of them may be replaced with a higher fitness one. The limit cases of this hypothesis are:
\begin{itemize}
\item the target is the model $\mathcal{X}_t = \mathcal{X}_m$: all neighbors have smaller fitness, so it is justifiable to replace some of them. The probability should be maximum: $r=1.0$.
\item the target is the one with the smallest fitness of the neighborhood $\mathcal{X}_t = \mathcal{X}_s$: any copy is already rewarding, so there is no urgency to replace any neighbor. The probability should be minimum: $r=0.0$.
\item Otherwise, the probability shoud be such as $0.0 < r < 1.0$.
\end{itemize}
This behavior is accomplished by the definition:
\begin{equation} 
    r_1 =  \frac{\phi_t - \phi_s}{\phi_m - \phi_s},  \label{woierjkjkj}
\end{equation} 
where $\phi_t = f(\mathcal{X}_t)$ is the fitness of the target agent $\mathcal{X}_t$, $\phi_s = f(\mathcal{X}_s)$ is the smallest fitness and $\phi_m = f(\mathcal{X}_m)$ is the maximum one (fitness of the model $\mathcal{X}_m$) on the influence neighborhood. 

We have tested other definitions for the probability and, surprinsingly, the following one gave the most interesting results:
\begin{equation} 
    r_2 = 1 -  \frac{\phi_t - \phi_s}{\phi_m - \phi_s}.  \label{coevo}
\end{equation} 
This is the opposite of the first attempt: $r=0.0$ if the target is the model and 1.0 if it has the smallest fitness. The two probabilities well be studied on the Results section.


\subsection{Random Network} \label{gfgfgfgfgfgfgfgf}

A network or a graph comprises a collection of vertices interconnected by a set of edges \cite{Newman2010,Mata2020}. Each edge, or connection, links a pair of vertices with the number of vertices defining the network size and the set of edges determining its connectivity. Vertices that share an edge are termed neighbors. Throughout this work the terms vertices, agents, and individuals have the same meaning. Various types of networks are used for solutions of physical and biological problems \cite{Vilela2020,Pineda2023,Silva2024}. In this work we will use the Erdös-Rényi (ER) network \cite{Solomonoff1951, Erdos1959} which has the following definition: consider a set of $M$ vertices and a parameter $\beta$ such that $0.0 < \beta < 1.0$, the probability of any two vertices being connected is $\beta$. This network has been used frequently as a first prototype to study the influence of the network topology on social dynamics \cite{Holme2006,Moore2015,Saeedian2019}.

The degree $k_i$ is the number of neighbors of node $i$ and the average degree $\mu$ of the network is defined as $\mu = (1/M) \sum_i^M k_i$ \cite{Gomes2024}. In the case of ER network, we have $\mu = \beta (M-1)$ \cite{Barabasi2016}. So we use $\mu$ as control parameter of the network instead of $\beta$. Figure \ref{figXsdff} shows a graphical representation of the ER network with two different values of $\mu$.

\section{Methods} \label{sec3}

A schematic illustration for our dynamic is presented in Figure \ref{fig0}. The input parameters are $M$, $L$, $p$, $N$, $K$ and $S$. The general algorithm is:
\begin{enumerate}
\item A sample of the random network is created and the time starts at $t=0$.
\item $t_g$ Monte Carlo Steps are executed until one agent finds the landscape maximum $F$. The computational cost is calculated (Eq. \ref{compcostlk}). 
\item Steps 1 and 2 constitute one time evolution. $S$ random time evolutions are perfomed and the average $\left\langle C \right\rangle$ computational cost is recorded.
\end{enumerate}
One Monte Carlo step is the execution of $M$ iterations. The algorithm for one iteration is:
\begin{itemize}
\item  One random node is selected to serve as the target agent.
\item The rewiring happens with probability of $r$ where $L$ links are removed and new $L$ links are created involving the target agent. 
\item The string update takes place and it can be either a mutation or a copy.
\end{itemize}
Fig. \ref{fig0}(a) illustrates this algorithm and \ref{fig0}(b) shows an example of the rewiring with $L=1$.

\begin{figure*} [h]
\centering
\includegraphics[width=3.0 in]{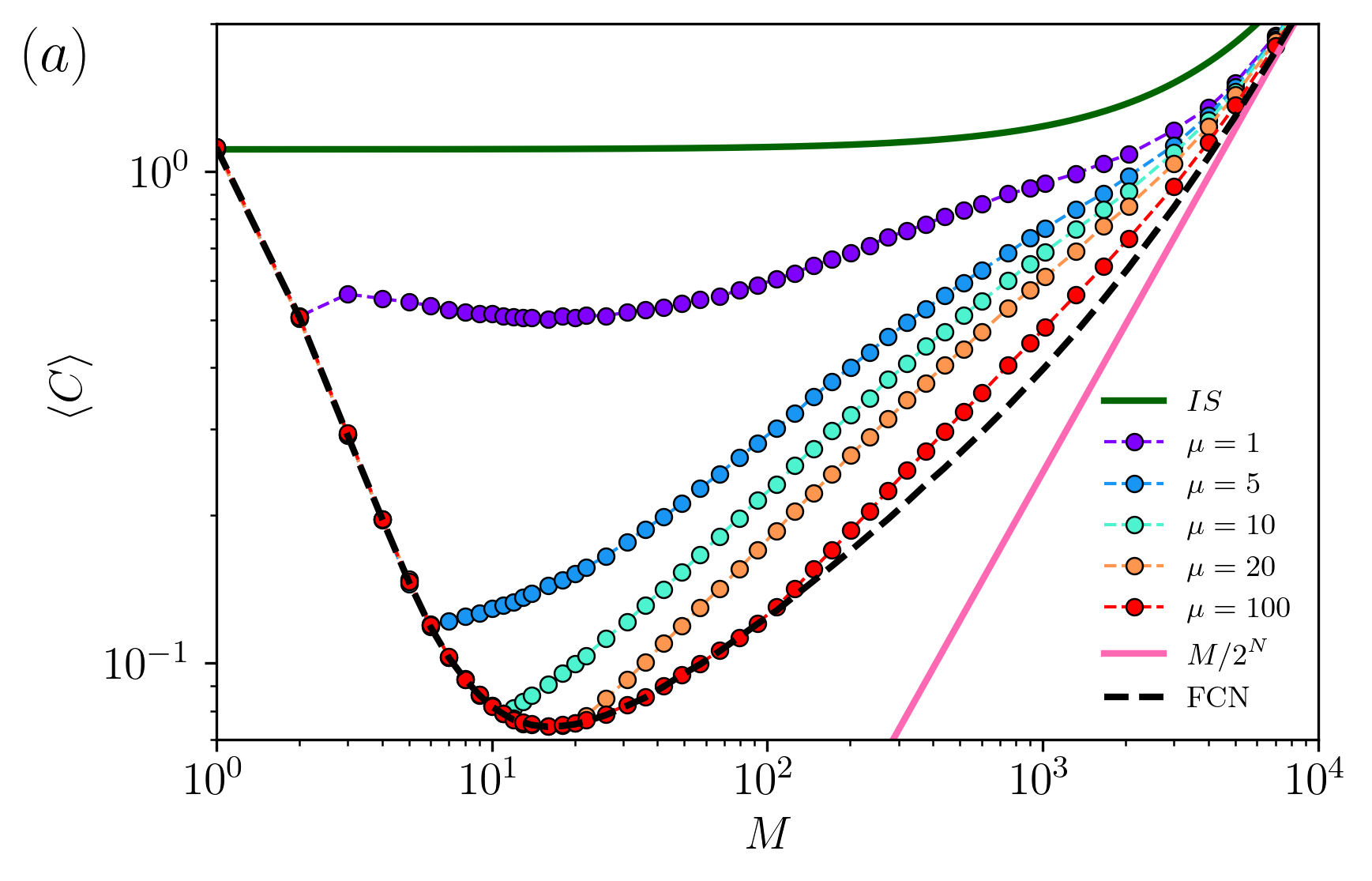}
\includegraphics[width=3.2 in]{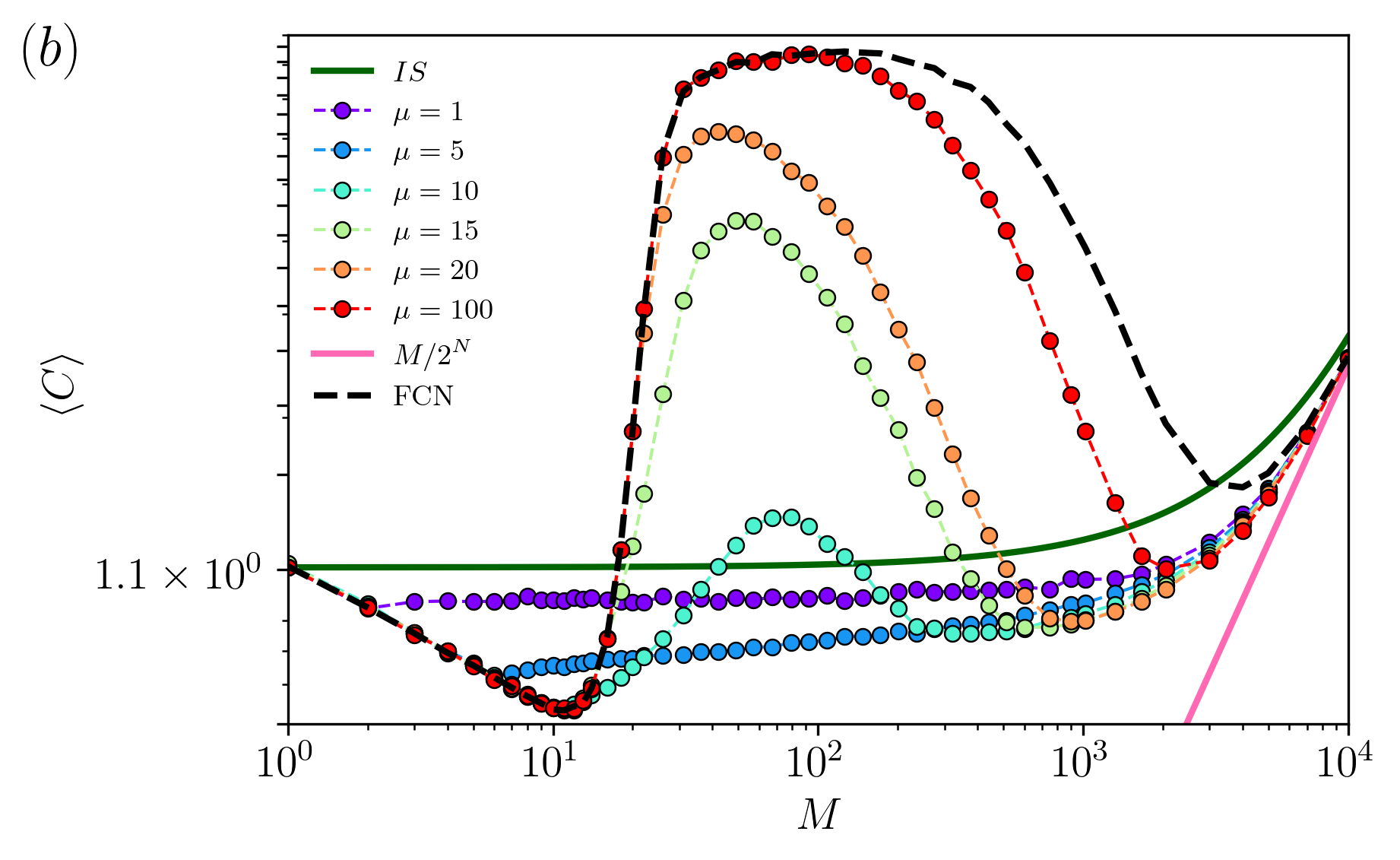}
\caption{Results of the Reference Case on the two investigated landscapes. Computational cost $\left\langle C\right\rangle$ as function of $M$ for different average degrees $\mu$ of the ERN with no coevolution. These results is the reference to which we will compare the results with coevolution to evaluate its impact. Parameter: $S=5\times 10^4$ samples. The green line is the IS (Eq. ref{indsearch}), black dashed line is the FCN result and the pink line is the approximation for large $M$. Purple, blue, green, orange and red circles correspond to an average degree of $\mu=1$, 5, 10, 20 and 100, respectively. (a) Easy landscape: $K=0$. (b) Difficult landscape: $K=4$.}
\label{fig1}
\end{figure*}

In this work we explored two values for the $K$ parameter: $K=0$ is the easy case and $K=4$ is the complicated one. We also investigated the following values for the coevolution: $L=0$, 1, 2, 3, 4, 5 and 6. We also fixed $N=12$ and the copy probability at $p=0.5$. At this value of $N$ we have $\lambda_{12} \approx 0.99978$. Since $(\lambda_{12})^M \approx e^{-M(1-\lambda_{12})}$, we can approximate the limit cases (low and large $M$) of the independent search from Eq. \ref{indsearch} as \cite{Fontanari2016a,Fontanari2015EPJB}:
\begin{equation}
\left\langle C_I \right\rangle \approx \left\{
\begin{array}{rl}
1.110, &  \text{ for } M <<   4545   \\
M/2^{12}, &  \text{ for } M >>  4545 
\end{array} 
 \right. \label{sekjeijkjdkjf}
\end{equation}

\subsection{Benchmark}

In order to better assess the impact of the coevolution porcess, we compare its results to 3 others data sets:
\begin{itemize}
    \item Independent Search (IS): when each agent independently searches for the global maximum. In this case there is no copy of the model strings, only mutation. Numerically, it can be achieved by setting $p=0$ or using a network with $k_i = 0$ for all agents $i$. Analytically, it can be calculated using Eq. \ref{indsearch}, and its limit cases for small and large $M$ are given by Eq. \ref{sekjeijkjdkjf}. This result will be displayed as a green solid line on the results.        
    \item Full Connected Network (FCN): when all agents are connected to all other agents: $k_i = M-1$ for all agents. This result will be displayed as a black dashed line on the results.
    \item Reference Case (RC): dynamics with a static network, no rewiring, equivalent to $L=0$. Coevolution will bring an advantage if it entails a lower computational cost $C$ than the Reference Case one.
\end{itemize}

The independent search is the simplest method, and the cheapest one if we think in terms of a company hiring employees to perform the task. The FCN case is also a simple one, however not the cheapest one. As $M$ increases, the cost to enable everybody to communicate with everybody also increases. Hence, for any particular dynamics to be called efficient, it must bring more efficiency than these two cases.

\subsection{Reference case}

Firstly we present the results with a static network: the Reference Case (RC). It is important to understand this case so we can highlight the effect of the coevolution. At $K=0$ (Figure \ref{fig1}(a)) the best result (lower $\left\langle C\right\rangle $) is given by the full connected network (FCN) and the worst by the independent search (IS). The increasing average degree $\mu$ interpolates from IS to the FCN case. As there is only one maximum (the global one) it becomes easier to increase the fitness with more neighbors to copy. However, in terms of system size $M$ there is one optimal value roughly between 15 and 20 which returns $\left\langle C\right\rangle \sim 0.07$. So, if there is too few neighbors (smaller $M$) or too much neighbors (larger $M$) the agents spend more time looking for the global maximum. At large $M \gtrsim 10^4$ the system behaviors as the limit case $M/2^N$.

\begin{figure*} [h]
\centering
\includegraphics[width=3.1 in]{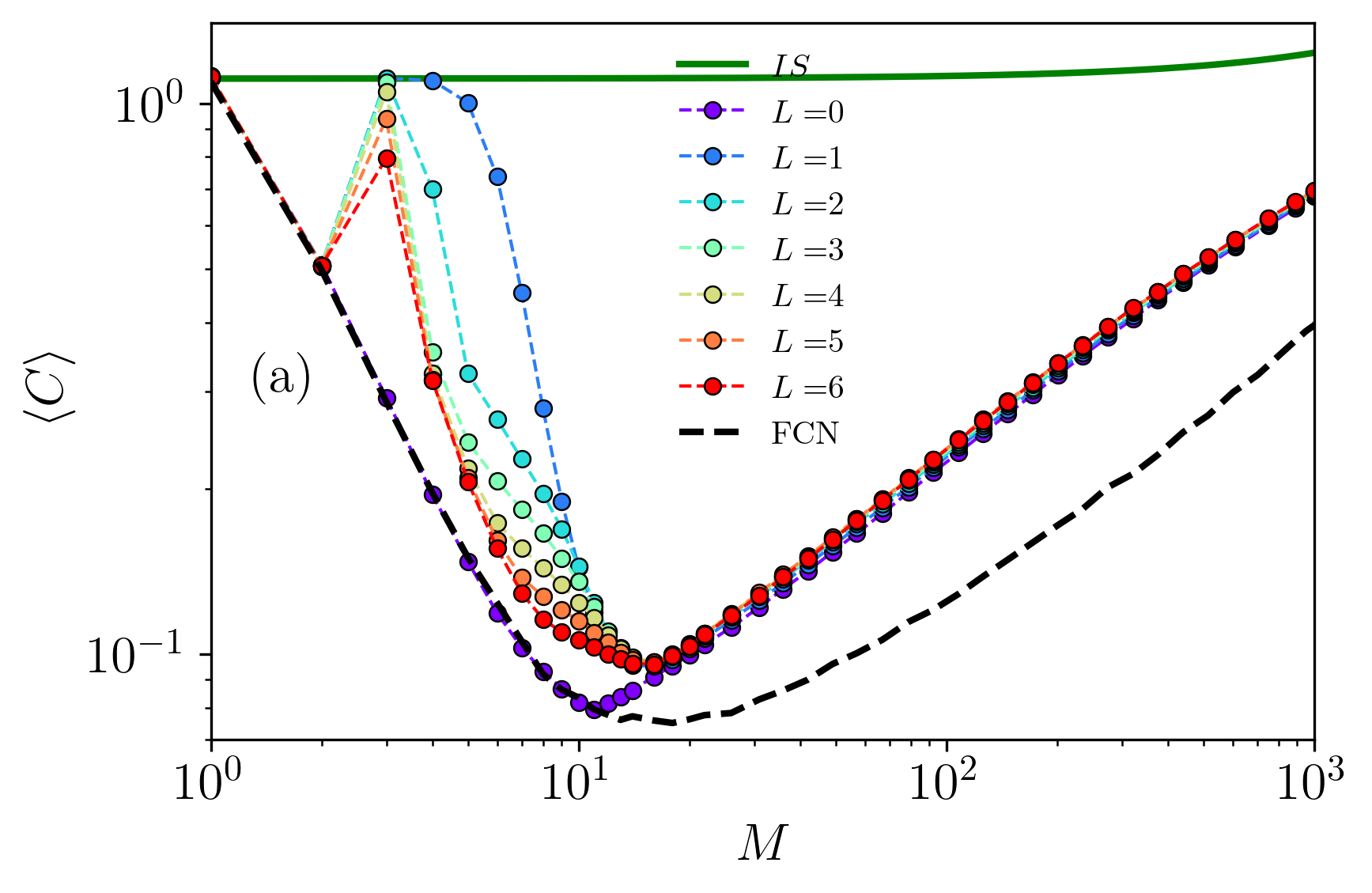}
\includegraphics[width=3.1 in]{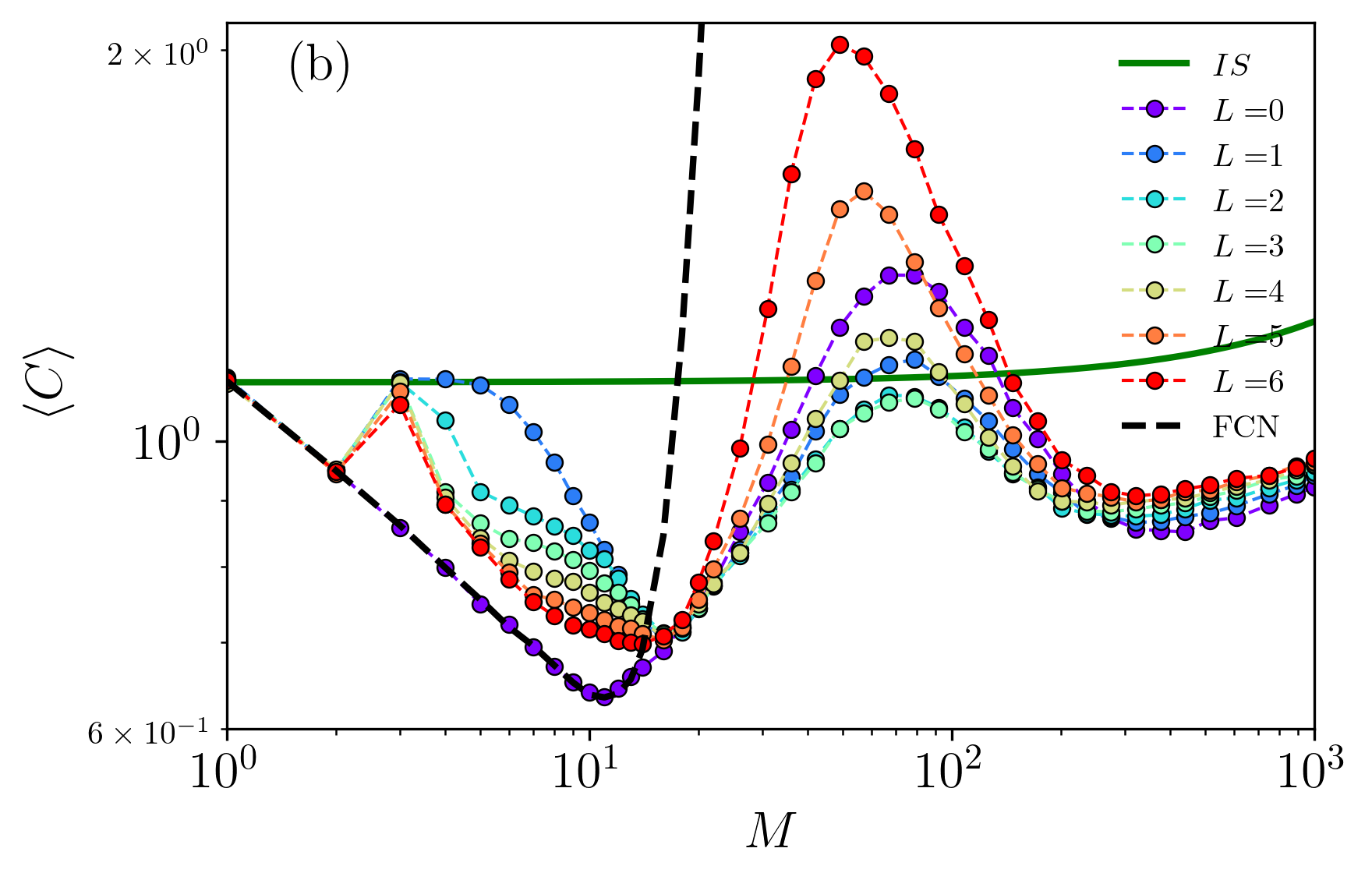}
\caption{Coevolution results using $r_1$ (Eq. \ref{woierjkjkj}) as rewiring probability on the two investigated landscapes. Computational cost $\left\langle C \right\rangle $ vs. $M$ for different values of modification $L$, limited by $L< k_i$ for each agent. $L=0$ means no coevolution. Parameters: $S=10^5$ samples and $\mu=10$. The green line is the IS (Eq. \ref{indsearch}) and the black dashed line is the FCN result. (a) Easy landscape: $K = 0$. (b) Difficult landscape: $K = 4$.}
\label{fig_CvsM_r1}
\end{figure*}

\begin{figure*} [h]
\centering
\includegraphics[width=3.1 in]{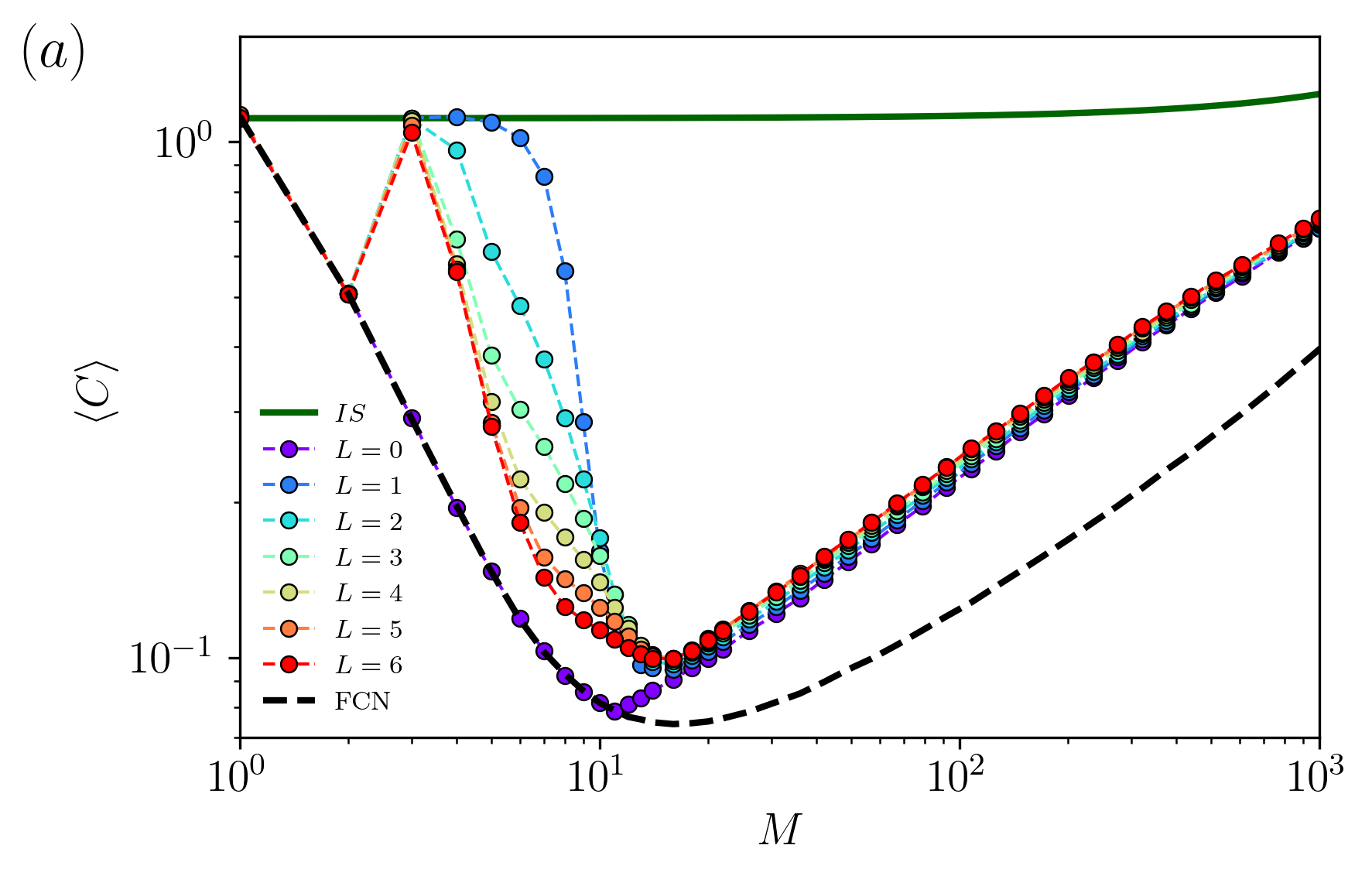}
\includegraphics[width=3.1 in]{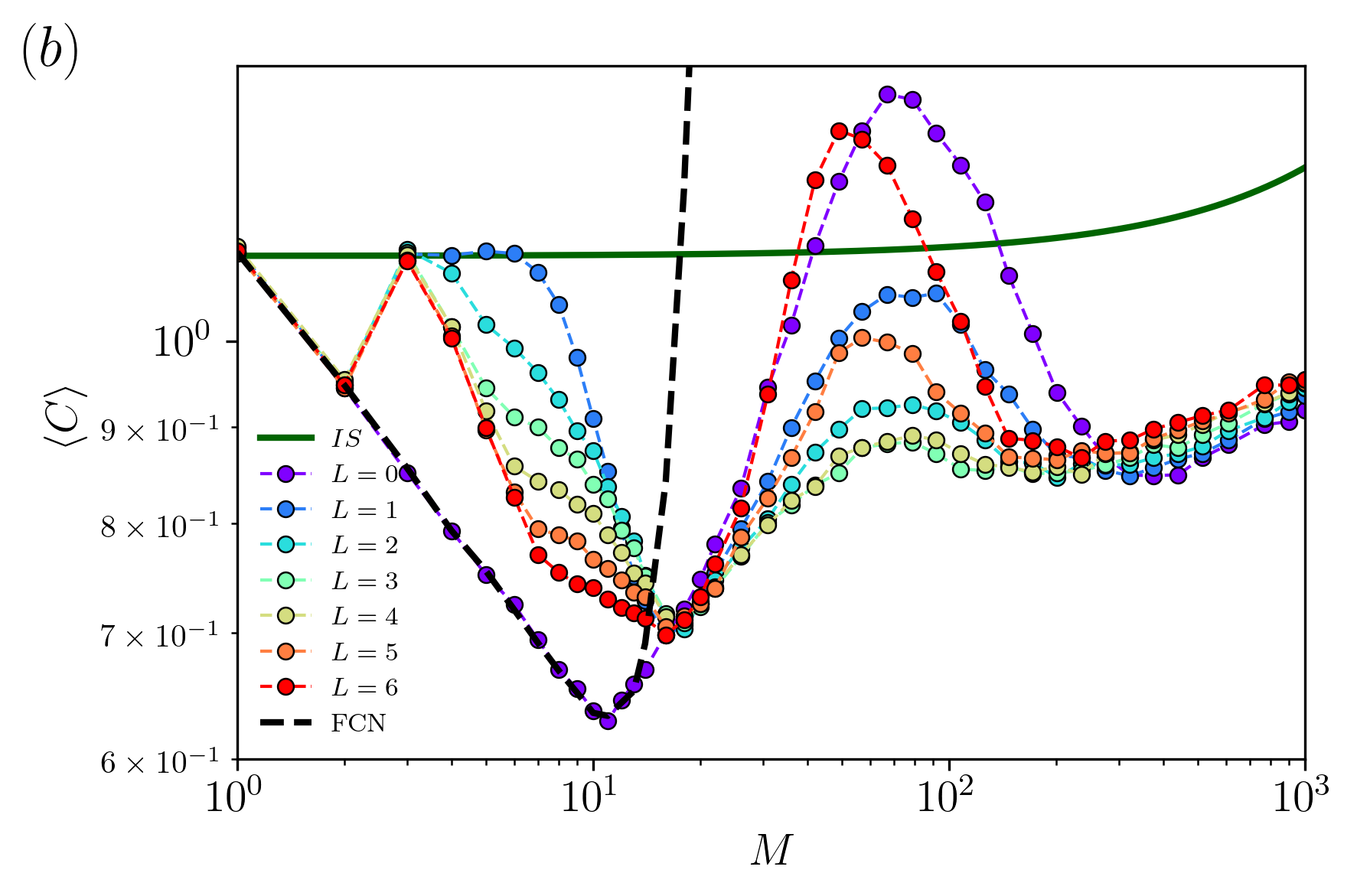}
\caption{Coevolution results using $r_2$ (Eq. \ref{coevo}) as rewiring probability on the two investigated landscapes. Computational cost $\left\langle C \right\rangle $ vs. $M$ for different values of modification $L$, limited by $L< k_i$ for each agent. $L=0$ means no coevolution. Parameters: $S=10^5$ and $\mu=10$. The green line is the IS (Eq. \ref{indsearch}) and the black dashed line is the FCN result. (a) Easy landscape: $K = 0$. (b) Difficult landscape: $K = 4$.}
\label{fig2b}
\end{figure*}

When $K=4$ (Figure \ref{fig1}(b)) the scenario is drastically different. As before there is an optimal value of $M$ roughly between 10 and 15. However there is a significant increase on the cost $\left\langle C\right\rangle $ at $M \gtrsim 20$ when $\mu$ is increased towards the FCN. As there are local maxima in this landscape, an agent can be trapped in one of these making it longer to find the global maxima. The results on these two scenarios resemble the similar case on the random geometric graph \cite{Gomes2019}.

\begin{figure*} 
 \centering
 \includegraphics[width=2.6 in]{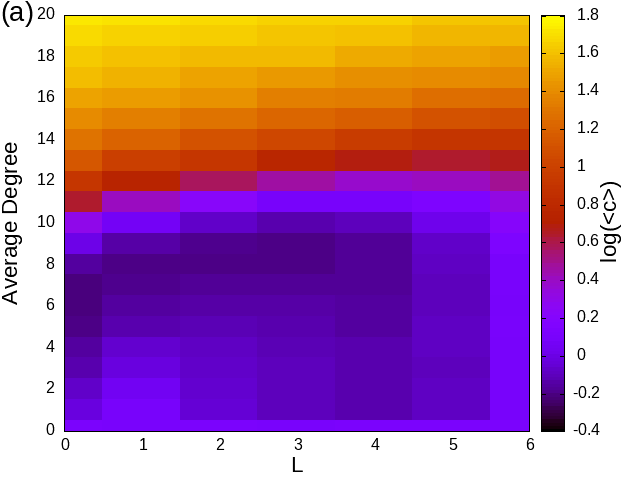}
 \includegraphics[width=2.6 in]{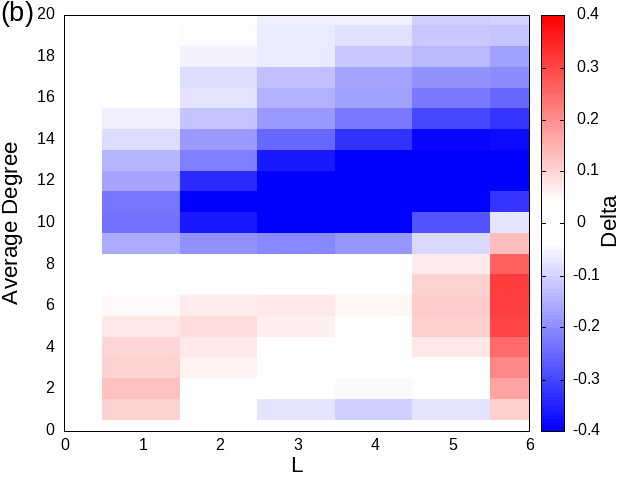}
 \caption{2D map of the computational cost $\left\langle C \right\rangle$ and its gain $\Delta$ in the $(L,\mu)$ space. Parameters: $K=4$, $M=67$ and $S=10^5$. (a) $\log \left\langle C \right\rangle$. (b) $\Delta$ defined on Eq. \ref{delta}. Positive values are indicated in red and negative values in blue.}
 \label{fig_ps}
\end{figure*}

\section{Results} \label{resultskj}

\subsection{Dynamics with coevolution}

To enhance the effect of coevolution, we restricted the values of the average degree network $\mu$ for the analysis. We observed different scenarios for different values of $\mu$: coevolution works for some cases and does not work at other cases. We say it works when the computational cost $\left\langle C \right\rangle $ is lower compared to the Reference Case, independent search (IS) and FCN case, keeping all other parameters the same. We will explore some local improvements on $\left\langle C \right\rangle $, at some values of $M$, at degree $\mu = 10.0$.

We begin with the results using $r_1$ (Eq. \ref{woierjkjkj}) as the rewiring probability, and Fig. \ref{fig_CvsM_r1}(a) shows the results for the ease case $K=0$. The results shows influence of $L$ only on small network sizes, around $M \lesssim 15$: the gain on $\left\langle C \right\rangle $ improves with $L$. However, at $M \geq 15$, $L$ has no influence at all. FCN result is still the best one for all $M$. At $K=4$ (Fig. \ref{fig_CvsM_r1}(b)) the results become more interesting. At small network size $M$, FCN is still the best result. However, at medium-sized networks ($20 \leq M \leq 120$), we see a significant improvement in computational cost with $L=2$ and $3$ around $M \sim 70$: $\left\langle C \right\rangle $ is slightly lower than the IS case. More interestingly, $\left\langle C \right\rangle $ rises back at $L=6$, greater than the $L=0$ value.

Moving on with the rewiring probability $r_2$ (Eq. \ref{coevo}), the results for the ease case $K=0$ (Fig. \ref{fig2b}(a)) are similar as the the ones with $r_1$. On the other hand, results for the difficult case $K=4$, depicted on Fig. \ref{fig2b}(b), are even more interesting. As before, there is an improvement on $\left\langle C \right\rangle $, but now the minimum is even lower: $\left\langle C \right\rangle \sim 0.9$. At higher $L$, $\left\langle C \right\rangle$ rises back, although lower than the $L=0$ value.

To better visualize the behavior of the $\left\langle C \right\rangle$ minimum at $M=67$, Fig. \ref{fig_ps}(a) displays a color map on the space $(L,\mu)$. However, a more interesting measure is the difference between the $L=3$ and $L=0$ values, defined as:
\begin{equation} \label{delta}
 \Delta  (L,\mu) = \left\langle C \right\rangle (L,\mu) - \left\langle C \right\rangle (0,\mu). 
\end{equation}
This quantity gives the improvement ($\Delta < 0$) on the cost $\left\langle C \right\rangle$ created by the coevolution, as can be seen on Fig. \ref{fig_ps}(b). The average degree has a great impact on this gain: at $L=6$ it goes from destroying it at $\mu = 6$ to improving the performance at $\mu = 12$. This result shows the tunable coevolution described here is a real tool to increase perfomance on the NK model.

\begin{figure} [h]
\centering
\includegraphics[width=3.2 in]{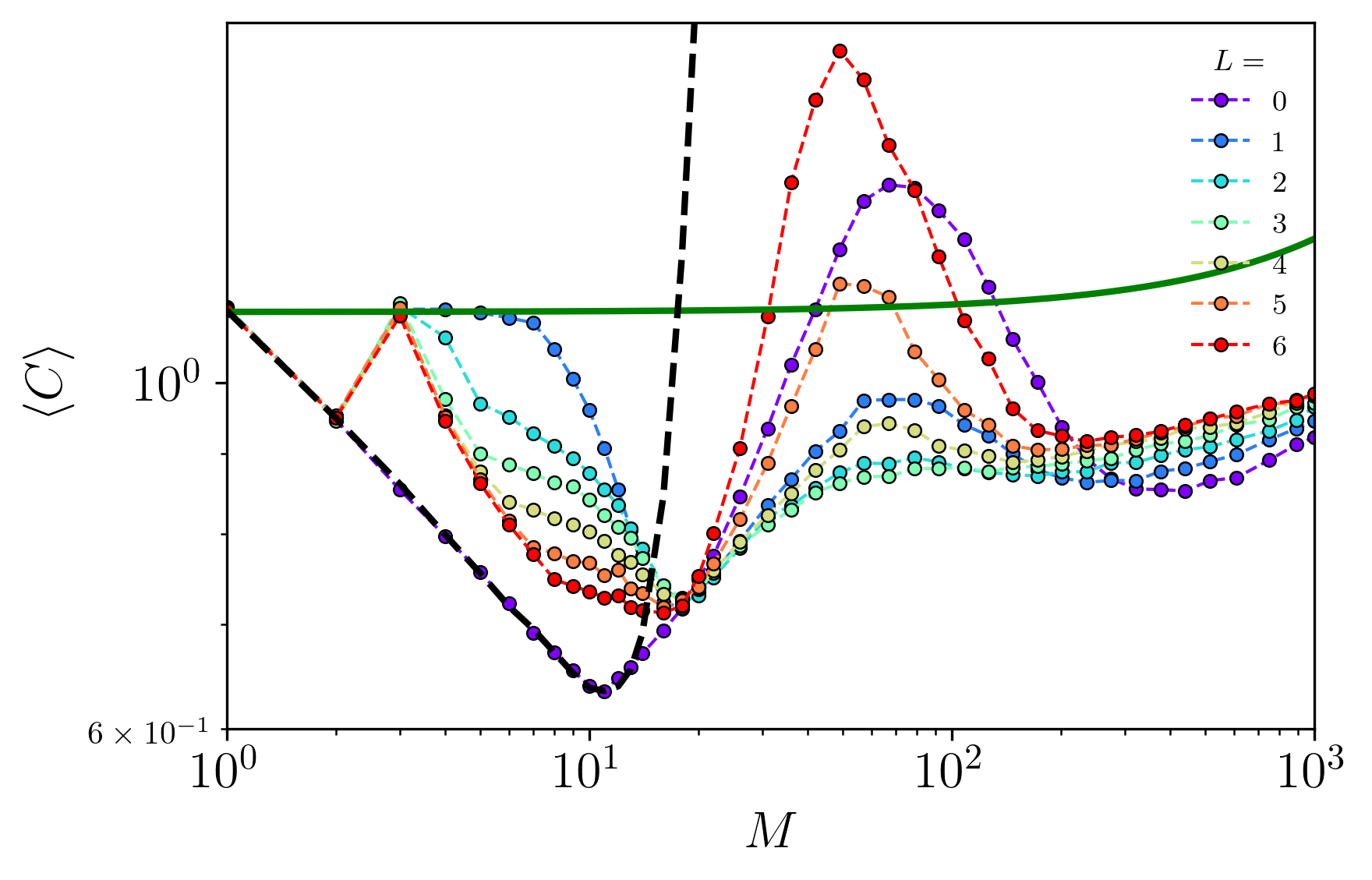}
\caption{Coevolution results using $r=1.0$ as rewiring probability (constant rewiring). Computational cost $\left\langle C \right\rangle $ vs. $M$ for different values of modification $L$, limited by $L< k_i$ for each agent. $L=0$ means no coevolution. Parameters: $S=10^5$, $\mu=10$ and $K = 4$. The green line is the IS (Eq. \ref{indsearch}) and the black dashed line is the FCN result.}
\label{fig_coevo_s66g14}
\end{figure}

\subsection{Rewiring probabilities}

\begin{figure*} [t]
\centering
\includegraphics[width=3.0 in]{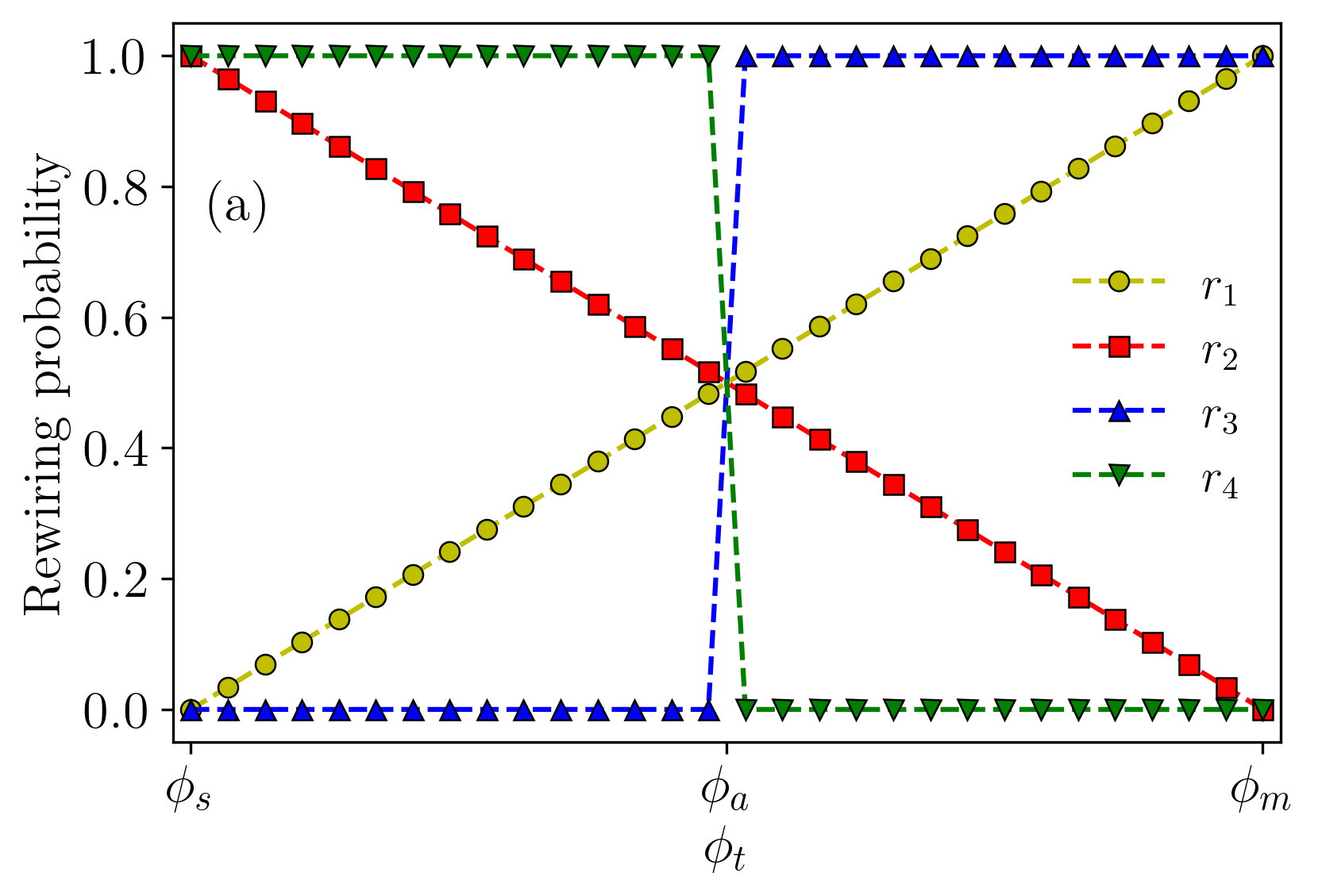}
\includegraphics[width=3.1 in]{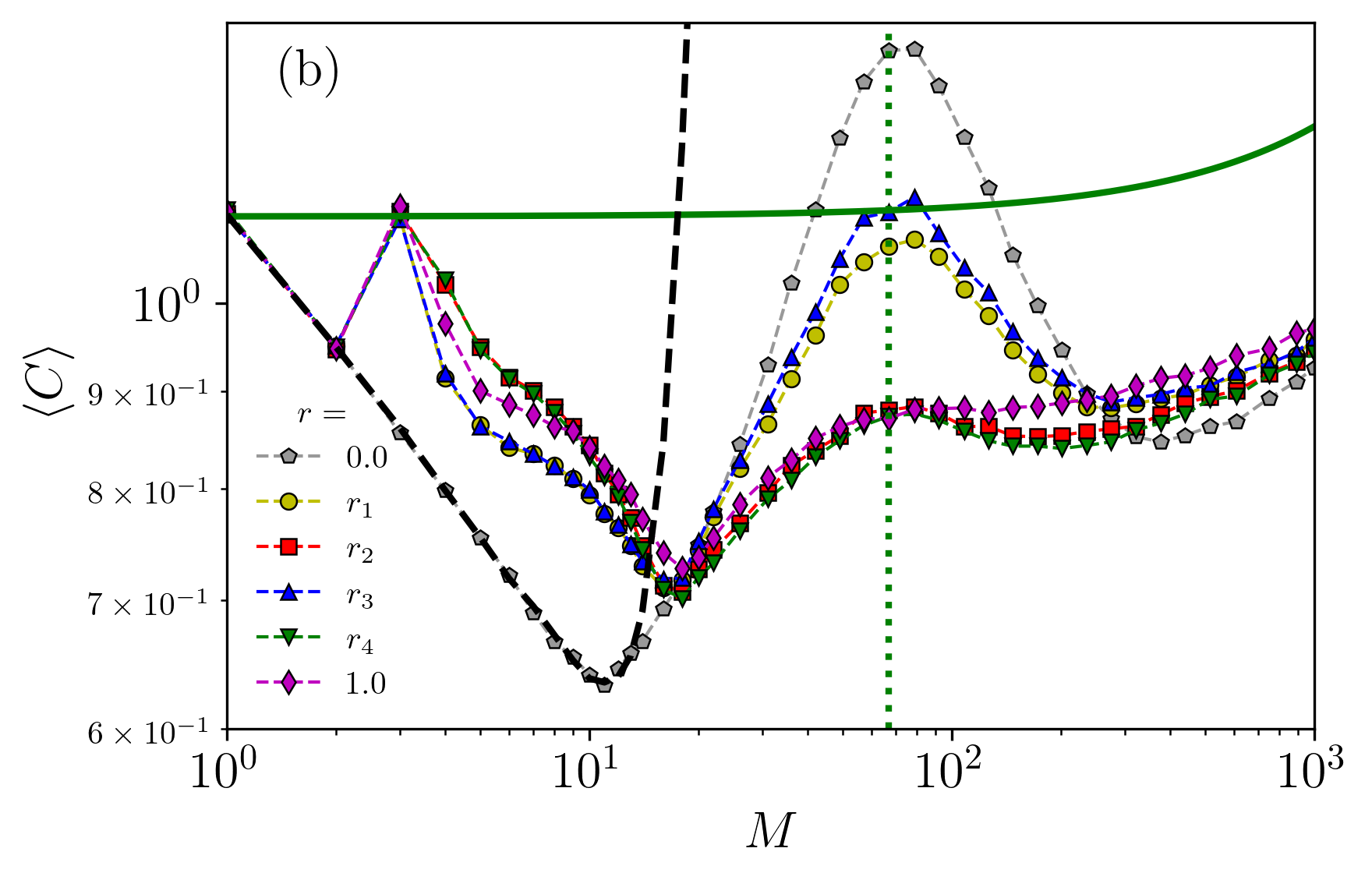}
\caption{Comparison of the results between different rewiring probabilities. (a) Behaviour of four different definitions: $r_1$ (Eq. \ref{woierjkjkj}), $r_2$ (Eq. \ref{coevo}), $r_3$ (Eq. \ref{defprobr3}) and $r_4$ (Eq. \ref{defprobr4}). (b) Computational cost $\left\langle C\right\rangle $ vs. $M$ for 6 possible rewiring probabilities at $L=3$. The black dashed line is the FCN result and the vertical dotted green line indicates the position $M = 67$. $r=0.0$ is the Reference case (no coevolution) and $r=1.0$ is the constant rewiring one. Parameters: $K=4$, $S=10^5$ and $\mu = 10$. The green line is the IS (Eq. \ref{indsearch}).}
\label{fig_probabilities}
\end{figure*}

It is clear from Figs. \ref{fig_CvsM_r1}(b) and \ref{fig2b}(b) that the best gain on $\left\langle C \right\rangle $ is achieved around $M \sim 67$ with $r=r_2$. It indicates that rewiring when $\phi_t$ is low is the key mechanism. To better understand this result, we studied other rewiring probabilities. Probability $r_1$ is maximum when the target is the model and $r_2$ is maximum when the target has the worst fitness on the neighborhood. The natural next step is the constant rewiring case $r=1.0$. Figure \ref{fig_coevo_s66g14} shows a similar result as the one using probability $r_2$ (compare with Fig. \ref{fig2b}(b)). Indeed, $r=1.0$ implies a constant rewiring when $\phi_t$ is low, similar to $r=r_2$, but it also includes constant rewiring when $\phi_t$ is higher. So, define two more rewiring probabilities:
\begin{equation}
r_3 = \left\{
\begin{array}{rl}
 0.0, &  \text{ if } \phi_t < \phi_a,    \\
 1.0, &  \text{ if } \phi_t \geq \phi_a,
\end{array} 
 \right.  \label{defprobr3}
\end{equation}
\begin{equation}
r_4 = \left\{
\begin{array}{rl}
 1.0, &  \text{ if } \phi_t < \phi_a,    \\
 0.0, &  \text{ if } \phi_t \geq \phi_a,
\end{array} 
 \right.  \label{defprobr4}
\end{equation}
where $\phi_a  = (\phi_s + \phi_m)/2$ is the average fitness between the limits of the neighborhood. The probabilities $r_1$ to $r_4$ are displayed on Fig. \ref{fig_probabilities}(a). Probability $r_3$ implies rewiring only when $\phi_t$ is closer to $\phi_m$ (larger than the average $\phi_a$). It is an extrapolation of probability $r_1$. On the other hand, probability $r_4$ implies rewiring only when $\phi_t$ is closer to the minimum $\phi_s$, an extrapolation of $r_2$.

\begin{figure} 
\centering
\includegraphics[width=3.0 in]{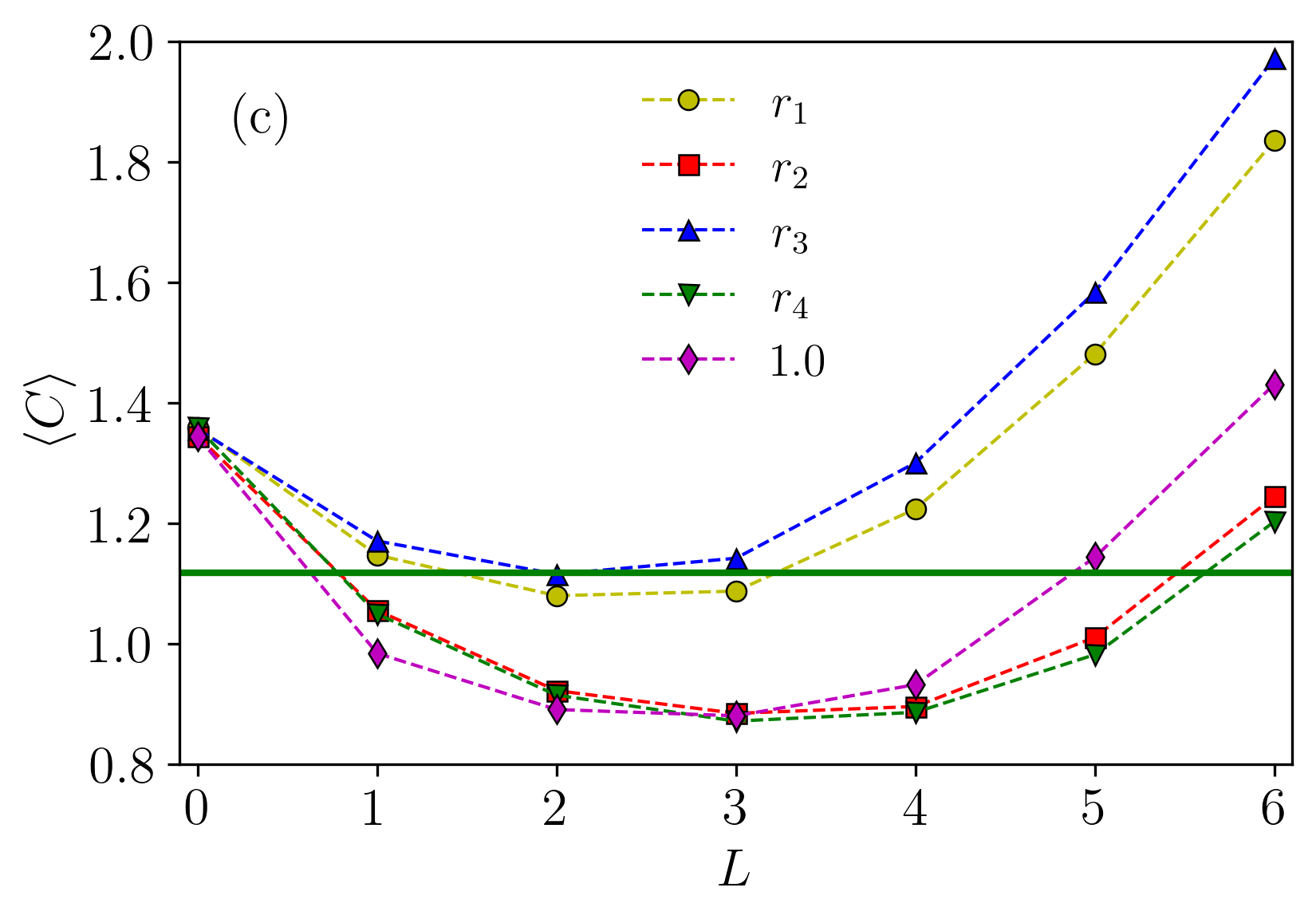}
\caption{Comparison of the results between different rewiring probabilities: Computational cost $\left\langle C\right\rangle $ vs. $L$. Parameters: $K=4$, $S=10^5$, $\mu = 10$ and $M=67$. The horizontal green line is the IS (Eq. \ref{indsearch}) value.}
\label{fig_CvsLsdf}
\end{figure}

Figure \ref{fig_probabilities}(b) shows that $r_4$ result is almost identical as the $r_2$ one, both very similar to the $r=1.0$ result. This is an strong evidence that rewiring when $\phi_t$ is low (lower than $\phi_a$) is the responsable mechanism for the observable gain on $\left\langle C\right\rangle $. Indeed, $r_3$ result is similar to the $r_1$ one, even slightly worst. The gain on the computational cost $\left\langle C\right\rangle $ can be better visualized on Fig. \ref{fig_CvsLsdf}: the best result (minimum on $\left\langle C\right\rangle $) is achieved with rewiring probabilities $r=r_2$, $r=r_4$ and $r=1.0$ at $L=3$.

\subsection{Discussion}

The goal of the rewiring is for the target to get a neighbor with higher fitness, worth being copied. The portion of agents with $\phi>\phi_t$ that can become target neighbor depends on $\phi_t$. If the target fitness is higher, $\phi_t > \phi_a$, close to $\phi_m$, there are less available agents with $\phi > \phi_t$. So, after a rewiring, the chance of the target to get a new neighbor with higher fitness may be reduced. The rewiring in this case can be ineffective. This is the case with rewiring probabilities $r=r_1$ and $r=r_3$. On the other hand, if the target has low fitness, $\phi_t < \phi_a$, there are more agents with $\phi > \phi_t$ available to become target neighbor. This can make the rewiring more effective when $\phi_t$ is low, so that the target has a reasonable chance to get acquainted with a higher fitness agent. This is the case with probabilities $r=r_2$, $r=r_4$ and $r=1.0$.

The described reasoning can explain why probabilities $r_1$ and $r_3$ do not bring any improvement on the computational cost, regardless the landscape. Additionally, probabilities $r_2$ and $r_4$ did not bring any improvement when $K=0$. On this ease landscape, any copy is advantageous because there are no local maxima. So, when the taget (with low $\phi_t$) makes a copy, the rewiring does not bring additional benefit. On the difficult landscape the picture is different. Probabilities $r_2$ and $r_4$ seems to make the difference at $L=3$ as the chance do get a new model as neighbor are relevant. At lower and higher $L$, this chance seems to be reduced. 

This scenario can explain why the constant rewiring $r=1.0$ has similar results as probabilities $r_2$ and $r_4$. It includes the high probability of rewiring for all values of $\phi_t$. If this fitness is low, close to $\phi_s$, the rewiring is worthwhile. If $\phi_t$ is high, close to $\phi_m$, the rewiring does not make any difference.

\section{Conclusions} \label{slkjeroiuejww}

We investigated the impact of coevolution on a cooperative process, where the target can substitute a few of his neighbors in order to solve the problem more rapidly. The neighbors are important because copy one of them (exploitation) is one possible strategy to improve the fitness. So, each agent analyses the fitness of his neighbors and uses this piece of information to decide wether or not to substitute a few of them. The number of neighbors to be replaced is an input parameter of the model. This creates a tunable mechanism enabling some sort of intensity on the coevolution process: a ``weak'' coevolution would be $L=1$, which has been studied on different social models. A ``stronger'' coevolution would be $L \geq 2$, which, to the best of our knowledge, has not been studied before.

We chose to apply this new coevolution model on a cooperative process. Using as prototype the NK model, our results showed a convoluted behavior of the system efficiency. A natural strategy for the target is to change some of his neighbors if they have worst (smaller) fitness than him. However, if the target fitness is high, there may not be enough agents with even higher fitness to be selected with reasonable probability. So, the rewiring in this case may not be useful. In the limit, when the target is the model, it is easier if no rewiring is done, so he continue to be the model. A selfish feeling seems to reward the overal system, akin to Adam Smith invisible hand metaphor. However, if the target has a low fitness, then it is easier to find new agents with higher fitness to be selected as a new neighbor. In this case, the rewiring may be a valid strategy. 

This work showed that this tunable coevolution can drastically change the behavior of the system efficiency. More studies can be interesting to study other details of this mechanism. Additionally, it opens some interesting perspectives such as the impact of this tunable coevolution on the phase diagram involving absorbing phase transition of models such as Axelrod's, voter model and others.


\section*{Acknowledgements}

This work received financial support from the brazillian agencies CNPq (project 405508/2021-2) and FAPEG (project no. 401425/2023-1). This research was also supported by LaMCAD/UFG. The authors thank the anonymous referees for the valuable comments.

\section*{Author contributions}

\textbf{Francis F. Franco:} Conceptualization, Methodology, Software, Formal analysis, Investigation, Writing - Original Draft, Writing - Review e Editing, \textbf{Paulo F. Gomes:} Conceptualization, Methodology, Formal analysis, Investigation, Writing - Original Draft, Writing - Review,  Editing, Supervision, Project administration. Authors declare that there are no competing interests.

\section*{Data Availability Statement}

The model was implemented using Fortran and the analysis and graphics were developed using Python (packages: Numpy \cite{Harris2020}, Matplotlib \cite{Hunter2007}, NetworkX \cite{Hagberg2008} and Pandas \cite{McKinney2010}) and Gnuplot.
All codes and data are available upon request. 

\appendix 

\section{Appendix}

%

\subsection{NK landscape} 

In this appendix we present a simpled description of how the function $f(\mathcal{X})$ is created for a given landscape. 

Following the formal definition from Eq. \ref{wkjeriuwlkj}, we have $f:\mathscr{D} \rightarrow \mathscr{R}$, where $\mathscr{D}$ is the set of $2^N$ binary vectors $\mathcal{X}$ and $\mathscr{R}$ is the real numbers set. The argument of the contributing functions $\Phi_i$ is given by Eq. \ref{lkjweoeiruhkjh}. To better understand this, suppose the vector $\mathcal{Y} =(y_0,y_1,y_2,y_3)= (0,1,1,0)$ on the landscape $N=4$ and $K=2$. There will be $K+1=3$ bits on the argument of each $\Phi_i$. The first bit from the i-th contribution is the i-th bit of $\mathcal{Y}$, while the other bits are the $K=2$ neighbors on the right. So, the first contribution of $\mathcal{Y}$ is $\Phi_0(y_0,y_1,y_2)=(0,1,1)$. The second contribution will be $\Phi_1(y_1,y_2,y_3)=\Phi_1(1,1,0)$. On the third contribution $\Phi_2$, the third bit will be the fifth one of $\mathcal{Y}$. However, the string $\mathcal{Y}$ has only $N=4$ bits, so the fifth bit will be the first one. This equivalent to subtract $N$ from the position of the bit: $5-4 = 1$. Thus: $\Phi_2(y_2,y_3,y_0)=\Phi_2(1,0,0)$. In the same way: $\Phi_3(y_3,y_0,y_1)=\Phi_2(0,0,1)$ . Putting together the four contributions, we have:
\begin{eqnarray}
f (0,1,1,0) &=& \tfrac{1}{4} \Phi_0(0,1,1) + \tfrac{1}{4} \Phi_1(1,1,0) \nonumber \\
 &+& \tfrac{1}{4} \Phi_2(1,0,0) + \tfrac{1}{4} \Phi_3(0,0,1).  \label{dkjeiurekjkdjkjd}
\end{eqnarray}
Other examples are:
\begin{eqnarray}
f (0,0,0,0) &=&  \tfrac{1}{4} \Phi_0(0,0,0) + \tfrac{1}{4} \Phi_1(0,0,0)  \nonumber \\
&+& \tfrac{1}{4} \Phi_2(0,0,0) + \tfrac{1}{4} \Phi_3(0,0,0) , \nonumber \\
f (0,0,1,0) &=& \tfrac{1}{4} \Phi_0(0,0,1) +\tfrac{1}{4} \Phi_1(0,1,0) \nonumber \\
&+& \tfrac{1}{4} \Phi_2(1,0,0) + \tfrac{1}{4} \Phi_3(0,0,0), \nonumber \\
f (1,0,1,0) &=& \tfrac{1}{4} \Phi_0(1,0,1) +\tfrac{1}{4} \Phi_1(0,1,0) \nonumber \\
&+& \tfrac{1}{4} \Phi_2(1,0,1) + \tfrac{1}{4} \Phi_3(0,1,0), \nonumber \\
f (0,1,1,1) &=& \tfrac{1}{4}  \Phi_0(0,1,1) + \tfrac{1}{4} \Phi_1(1,1,1) \nonumber \\
&+& \tfrac{1}{4} \Phi_2(1,1,0) + \tfrac{1}{4} \Phi_3(1,0,1) , \nonumber \\
f (1,1,1,1) &=& \tfrac{1}{4} \Phi_0(1,1,1) + \tfrac{1}{4} \Phi_1(1,1,1)  \nonumber \\
&+& \tfrac{1}{4} \Phi_2(1,1,1) + \tfrac{1}{4} \Phi_3(1,1,1). \nonumber 
\end{eqnarray}

It is worth mention that each argument on the functions $\Phi_i$ contains $K+1$ bits. So, there are $2^{K+1}$ possible combinations for the $N$ contributions from each vector.

\subsection{Base 2 and 10}

Conceptually, the description from previous section is enough. However, computationally, it is compelling to represent the vectors $\mathcal{X}$ and the arguments of the contributions $\Phi_i$ as integers. Both of them are binary sequences, 0s and 1s, numbers in base 2, and thus, can be converted to decimal numbers (base 10). In general, considering $N=4$: $(x_0,x_1,x_2,x_3) = x_3 \cdot 2^0 + x_2 \cdot 2^1 + x_1 \cdot 2^2 + x_0 \cdot 2^3$, starting the counting from the right. Thus,
\begin{eqnarray}
(0,0,0,0) = 0, \qquad (0,0,0,1) = 1, \nonumber \\
(0,0,1,0) = 2, \qquad (0,1,0,0) = 4, \nonumber \\
(1,0,0,0) = 8, \qquad (0,1,0,1) = 5. \nonumber
\end{eqnarray}
Computationally, this is much more efficient because we can record each string by its base 10 correspondent number (an integer), which takes much less memory than the string with $N$ bits.

We can do the same for the arguments of the contribution functions. Eq. \ref{dkjeiurekjkdjkjd} would be:
\begin{eqnarray}
f (0,1,1,0) &=& f(6) , \nonumber \\
&=& \frac{1}{4} \left[  \Phi_0(3) +  \Phi_1(6) +  \Phi_2(4) +  \Phi_3(1) \right]. \nonumber
\end{eqnarray}
Other example: the fitness of the vector $\mathcal{X} = 7$ considering $N=4$ and $K=3$ is:
\begin{eqnarray}
f (7) &=& f (0,1,1,1), \nonumber \\
&=& \tfrac{1}{4}  \Phi_0(0,1,1) + \tfrac{1}{4} \Phi_1(1,1,1) \nonumber \\
&+& \tfrac{1}{4} \Phi_2(1,1,0) + \tfrac{1}{4} \Phi_3(1,0,1), \nonumber \\
 &=& \frac{1}{4} \left[ \Phi_0(3) + \Phi_1(7) + \Phi_2(6) + \Phi_3(5) \right]. \nonumber 
\end{eqnarray}

This binary to decimal conversion is performed only once, when the landscaped is built, in the beginning of the algorithm. The fitness is the function $f(i)$ where $i=0,1,2,3,...,2^N-1$. Likewise, the contributions are functions $\Phi_i(j)$ where $i=0,1,2,3,...,N-1$ and $j=0,1,2,3,...,2^{K+1}-1$. Initially, the vectors $\mathcal{X}$ are randomly generated, which can be done by generating random integers from 0 to $2^N-1$ using a uniform distribution.

Once the arguments are defined, we have to define the functions $\Phi_i(j)$ itself. In this work, we use the NK landscape for a cooperative problem where the simulation comes to a halt when one agent finds the global maximum. So, it is interesting to keep the fitness on the range $\left[ 0,1\right]$. Thus, we define the functions $\Phi_i(j)$ as a random number from the uniform distribution on the same range $\left[ 0,1\right]$.

\end{document}